\documentclass[letter]{aa} % for the letters 
\usepackage{graphicx}
\usepackage{lscape}
\usepackage{amsmath}
\usepackage{txfonts}
\usepackage{url}
\usepackage{natbib}
\usepackage{xcolor}

\begin{document}

\title{Yellow supergiants as supernova progenitors: an indication of strong mass loss for red supergiants?}

\author{Cyril Georgy}%\inst{\ref{inst1}}}

\authorrunning{Cyril Georgy}

\institute{Centre de Recherche Astrophysique de Lyon, Ecole Normale Sup\'erieure de Lyon, 46, all\'ee d'Italie, F-69384 Lyon cedex 07, France\label{inst1}}

\date{Received ; accepted }

% RESUME %%%%%%%%%%%%%%%%%%%%%%%%%%%%%%%%%
\abstract
  % context heading (optional) leave it empty if necessary  
{The increasing number of observed supernova events allows for finding ever more frequently the progenitor star in archive images. In a few cases, the progenitor star is a yellow supergiant star. The estimated position in the Hertzsprung-Russell diagram of these stars is not compatible with the theoretical tracks of classical single-star models.}
{According to several authors, the mass-loss rates during the red supergiant phase could be underestimated. We study the impact of an increase in these mass-loss rates on the position of $12$ to $15\, M_{\sun}$ stars at the end of their nuclear lives, in order to reconcile the theoretical tracks with the observed yellow supergiant progenitors.}
{We have performed calculations of $12$ to $15\, M_{\sun}$ rotating stellar models using the Geneva stellar evolution code. To account for the uncertainties in the mass-loss rates during the RSG phase, we increased the mass-loss rate of the star (between $3$ and $10$ times the standard one) during that phase and compared the evolution of stars undergoing such high mass-loss rates with models computed with the standard mass-loss prescription.}
{We show that the final position of the models in the Hertzsprung-Russell diagram depends on the mass loss they undergo during the red supergiant phase. With an increased mass-loss rate, we find that some models end their nuclear life at positions that are compatible with the observed position of several supernova progenitors. We conclude that an increased mass-loss rate (whose physical mechanism still needs to be clarified) allows single-star models to simultaneously reproduce the estimated position in the HRD of the YSG SN progenitors, as well as the SN type.}
{}

\keywords{stars: evolution -- stars: rotation -- stars: massive -- stars: mass-loss -- stars: supergiants}

\maketitle
%%%%%%%%%%%%%%%%%%%%%%%%%%%%%%%%%%%%%%%%

% INTRODUCTION %%%%%%%%%%%%%%%%%%%%%%%%%%%%%%
\section{Introduction}\label{SecIntro}

During the past decade, the increase in the detection number of supernova (SN) events, combined with the growing number of identifications of the progenitor star in archive images \citep{Li2007a,Smartt2009b}, become an important test for stellar evolutions models. SNe are classified in several categories, based on their light curve and spectra \citep{Filippenko1997a}. The two main groups are the type II SNe (exhibiting hydrogen lines in their spectrum) and type I (without such lines). Most core collapse SNe are type IIP SNe \citep{Smartt2009a,Smith2011b}, requiring a progenitor with massive hydrogen envelope that produces the typical ``plateau'' in the light curve. The plateau is due to the photosphere moving inwards through the mass of the ejecta while the ejecta expands, leading to a constant radius photosphere and a plateau in the lightcurve. Fewer core-collapse SNe are found to be type IIL or type IIb, whose progenitor is thought to have a much less massive hydrogen envelope. These two  types of SNe require progenitors that encounter stronger than usual mass loss to partially remove their hydrogen-rich layers \citep{Heger2003a,Eldridge2004a}. Finally the type I core collapse supernovae are separated into two subtypes, type Ib (with helium features in the spectrum) and Ic (without helium).

Among recent observations of supernova progenitors there are some cases that are challenging to explain in the frame of the classical single-star evolution. The progenitor of SN 2008ax \citep{Crockett2008a} lies in the Hertzsprung-Russell diagram (HRD) between the two main areas where the classical non-rotating stellar model tracks end: the red supergiants (RSG) branch, on the one hand (with final $\log(T_\text{eff}/\text{K}) \sim 3.55$), and the Wolf-Rayet (WR) location, on the other (with final $\log(T_\text{eff}/\text{K})\sim 4.5 - 5$). However, the position of the progenitor star can be explained by a single star leading to a WN star at the end of the evolution \citep{Crockett2008a}, and recent theoretical tracks with rotation and increased mass-loss rate allow reproduction of the estimated position of its progenitor \citep{Ekstrom2011a}. Several other SNe arise from stars that are not RSG: type IIb SN 1993J  \citep{Aldering1994a,Maund2004a}, type IIP SN 2008cn \citep{Elias-Rosa2009a}, type IIL SN 2009kr \citep{Fraser2010a,Elias-Rosa2010a}, and type IIb SN 2011dh \citep{Maund2011a}, all have a yellow supergiant (YSG) progenitors\footnote{\footnotesize{The progenitor of SN 1993 is classified as a K supergiant \citep{Maund2004a}. However, its $\log(T_\text{eff}/\text{K}) = 3.63$ makes it a ``hot'' RSG instead of a YSG.}}, with surface effective temperature in the range $3.6 \lesssim \log\left( T_\text{eff}/K\right) \lesssim 3.8$. For most of these progenitors, the observational data could be compatible with a binary system, which could explain the unusual position of these stars in the HRD just before their explosion \citep{Podsiadlowski1993a,Elias-Rosa2009a}. This is particularly true for SN 1993J, which is most probably a binary, and where strong observational constraints exist about the secondary, and where the progenitor can be well explained by a binary stellar model \citep[see][and references therein]{Stancliffe2009a}. However, the recent observation of SN 2011dh seems to indicate a single-star progenitor \citep{Maund2011a}. Single-star evolution models should thus be able to explain at least this last peculiar case, both the SN type and the estimated position in the HRD of the progenitor.

% FIGURE HRD &&&&&&
\begin{figure*}
\centering
\includegraphics[width=.45\textwidth]{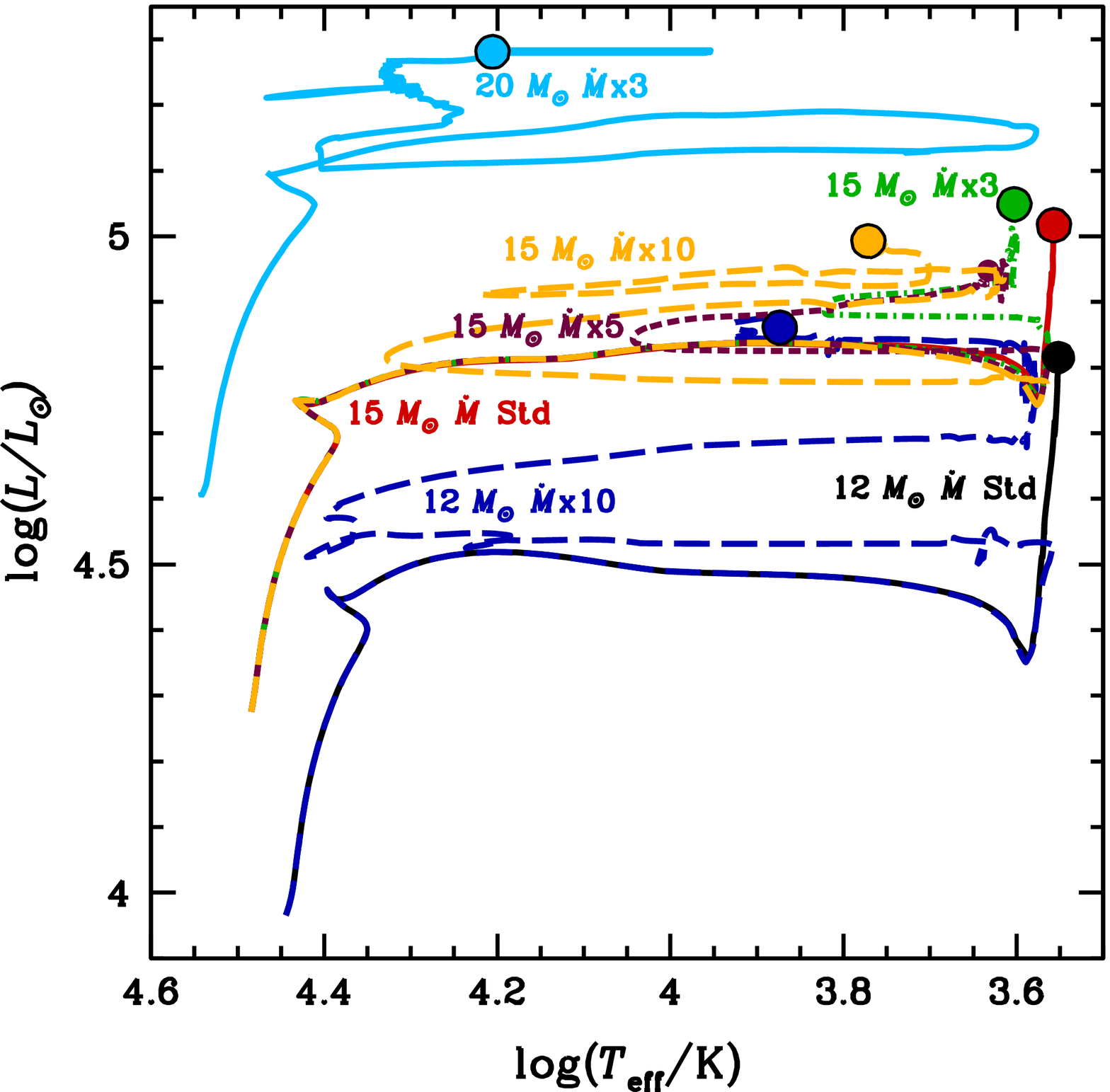}\hspace{.5cm}\includegraphics[width=.45\textwidth]{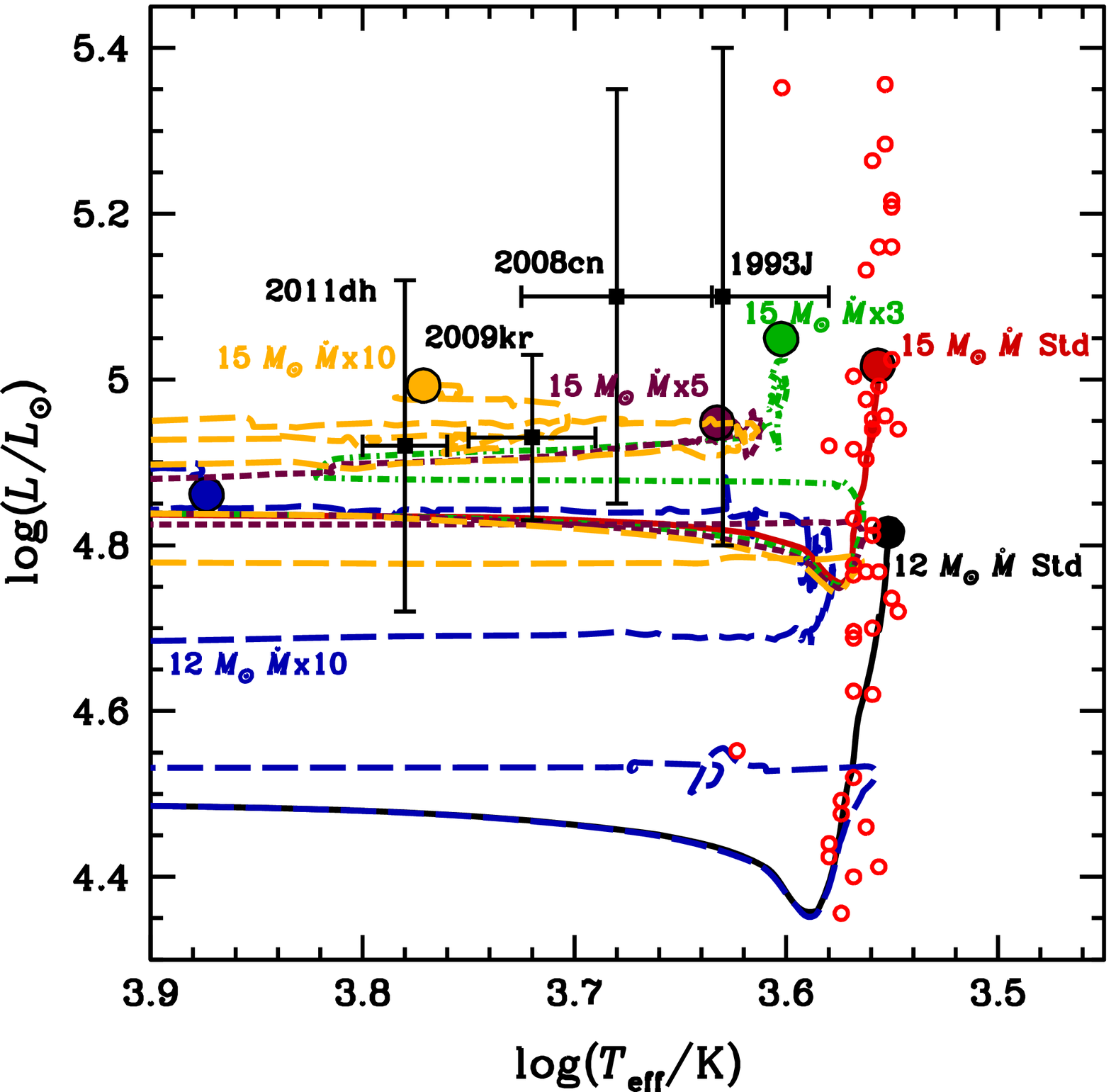}
\caption{\textit{Left panel:} HRD for our set of models. The initial mass and mass-loss prescription is indicated near the tracks. The circles indicate the end points of the simulations. \textit{Right panel:} Zoom on the region where the end points of the tracks are situated. The observational data are from \citet{Maund2004a}, \citet{Elias-Rosa2009a,Elias-Rosa2010a}, and \citet{Maund2011a}. The small red points represent the observed positions of Galactic RSG \citep{Levesque2005a}.}
\label{FigHRD}
\end{figure*}

One of the most important physical process governing the evolution of massive stars is mass loss. Combined with rotation, it changes the evolution of the star radically \citep[see the review by][]{Maeder2011a}. However, the mass-loss rates inferred observationally or theoretically remain uncertain. One of the most popular mass-loss prescriptions for the RSG phase is given by \citet{deJager1988a}, and has been confirmed observationally \citep{Crowther2001a,Mauron2011a}. On the other hand, some authors claim that the mass-loss rates during the RSG phase could have been underestimated \citep{vanLoon2005a,Davies2008a,Smith2009a,Moriya2011a}. Several authors have already studied the effects of such an increased mass-loss rate on the evolution of non-rotating single stars. \citet{Vanbeveren1998a,Vanbeveren1998b} show that the RSG and WR lifetimes critically depend on the mass-loss prescription used during the RSG phase. \citet{Eldridge2004a} studied particularly the effects on the remnant mass and the final mass and ejected mass. They show that an increased mass-loss rate leads to lower remnant mass, and smaller hydrogen content in the ejecta.

In \citet{Ekstrom2011a}, an increased mass-loss rate during the RSG phase was already used for stars with an initial mass $M \ge 20\, M_{\sun}$ during the RSG phase. In this letter, we extend this prescription to lower initial mass models, and explore systematically the consequences of an increased mass-loss rate on the final position of the model in the HRD, as well as on the SN type produced at the end of the stellar life. In Sect.~\ref{SecPhysics}, we briefly recall the physics included in our models, particularly the mass-loss prescriptions used. In Sect.~\ref{SecEvolution}, we describe the effect of the increased mass-loss rate on the evolutionary tracks in the HRD. In Sect.~\ref{SecCompa}, we discuss the consequences on the supernova progenitors and compare them with some recent observations. We present our conclusions in Sect.~\ref{SecConclu}.

% SECTION 1 %%%%%%%%%%%%%%%%%%%%%%%%%%%%%%
\section{Physics of the models}\label{SecPhysics}

The models computed for this work contain exactly the same physical ingredients as the models presented in \citet{Ekstrom2011a}. We mention here the main points of interest for this work:
\begin{itemize}
\item The initial abundances are $X=0.720$, $Y=0.266$, and $Z=0.014$, with a solar mixture for the heavy elements \citep[][and \citealt{Cunha2006a} for the Ne abundance]{Asplund2005a}. This corresponds to the initial composition of the Sun.
\item The models are computed with the inclusion of the effects of the rotation, according to the theoretical developments of \citet{Zahn1992a} and \citet{Maeder1998a}. Vertical shear and meridional circulation are included in this approach.
\item In the mass domain studied in this paper, the mass-loss rates are taken from \citet{deJager1988a} for stars with $\log(T_\text{eff}/\text{K}) > 3.7$. For stars with $\log(T_\text{eff}/\text{K}) \le 3.7$, we use a linear fit from the data of \citet{Sylvester1998a} and \citet{vanLoon1999a} \citep[see][]{Crowther2001a}.
\item Due to the uncertainty on the mass-loss rate in the low-$T_\text{eff}$ part of the HRD (see Sect.~\ref{SecIntro}), we apply several prescriptions in this paper. We apply a multiplicative factor between three and ten to the standard mass-loss rate prescription described above, when the effective temperature of the star is lower than $\log(T_\text{eff}/\text{K}) \le 3.8$. The detailed list of the computed models is shown below (see Table~\ref{TabListModels}). The exact physical mechanism of this increased mass loss still has to be determined. One of the candidates for such a mechanism is the pulsation-driven winds \citep{Heger1997a,Yoon2010b}. Binary interactions can also lead to an increased mass loss during the RSG phase \citep{vanBeveren2007a}.
\end{itemize}

To study the impact of an increased mass-loss rate during the RSG phase, we computed rotating stellar models of $12$ and $15\, M_{\sun}$, with an initial equatorial velocity given by $v/v_\text{crit} = 0.4$ \citep[which reproduces the observed mean velocity of B stars well during the main sequence (MS), see][]{Ekstrom2011a}. We performed computations with the standard and increased mass-loss rates. The list of the computed models is shown in Table~\ref{TabListModels}. Most of the models were computed up to the end of central carbon burning. After that point, the central evolution of the star is supposed to be so quick that the surface properties ($L$, $T_\text{eff}$) are no longer modified. However, to verify this assessment, we computed two models up to the end of central neon burning (models 15Std and 15x10). After this the star only has a few years before core collapse so the observable nature of the star will change only slightly. Luminosity and effective temperature of models computed up to the end of central Ne-burning only changed by less than 0.004 dex. Only one model, 15x5, was computed only up to the end of the central helium burning, and its final position should be considered a lower limit to the surface temperature and luminosity.

\begin{table}
\caption{List of the computed models.}
\centering
\begin{tabular}{ccc||ccc}
\hline\hline
label & $M_\text{ini}$ & $\dot{M}$ &label & $M_\text{ini}$ & $\dot{M}$ \\
 & $[M_{\sun}]$ & $[M_{\sun}\cdot\text{yr}^{-1}]$ & & $[M_{\sun}]$ & $[M_{\sun}\cdot\text{yr}^{-1}]$\\
\hline
12Std & $12$ & Std\tablefootmark{1} & 15x3 & $15$ & $\text{Std}\cdot 3$ \\
12x10 & $12$ & $\text{Std}\cdot 10$ & 15x5 & $15$ & $\text{Std}\cdot 5$ \\
15Std & $15$ & Std\tablefootmark{1} & 15x10 & $15$ & $\text{Std}\cdot 10$ \\
\hline
\hline
\end{tabular}
\tablefoot{\tablefoottext{1}{See text.}}
\label{TabListModels}
\end{table}

% SECTION 2 %%%%%%%%%%%%%%%%%%%%%%%%%%%%%%
\section{Impact on the stellar evolution}\label{SecEvolution}

In the left-hand panel of Fig.~\ref{FigHRD}, we show the evolutionary tracks of our models in the HRD. As a comparison, we also plot the track of the rotating $20\, M_{\sun}$ model of \citet{Ekstrom2011a}, which was also calculated with a mass loss increased by a factor of three during the RSG phase. The models computed with the standard mass-loss prescription (black and red tracks for the $12$ and $15\, M_{\sun}$ models, respectively) follow a classical evolutionary path, crossing the HRD after the MS, and then climbing the RSG branch at roughly constant $T_\text{eff}$. The models with increased mass-loss rates (green, purple, and orange tracks for the mass-loss rates increased by factors of $3$, $5$, and $10$, respectively) show the same behaviour. After they start to climb the supergiant branch, they all become hotter, going back in the bluer part of the HRD in a blue loop. They then come back in the yellow supergiant area, where they end their lives. We also see that the bluewards extension of the loop depends on the mass-loss rates: the higher the mass-loss rate, the hotter the star becomes during the loop (up to $\log(T_\text{eff}/\text{K}) > 4$).

% FIGURE Mdot &&&&&&
\begin{figure}
\centering
\includegraphics[width=.45\textwidth]{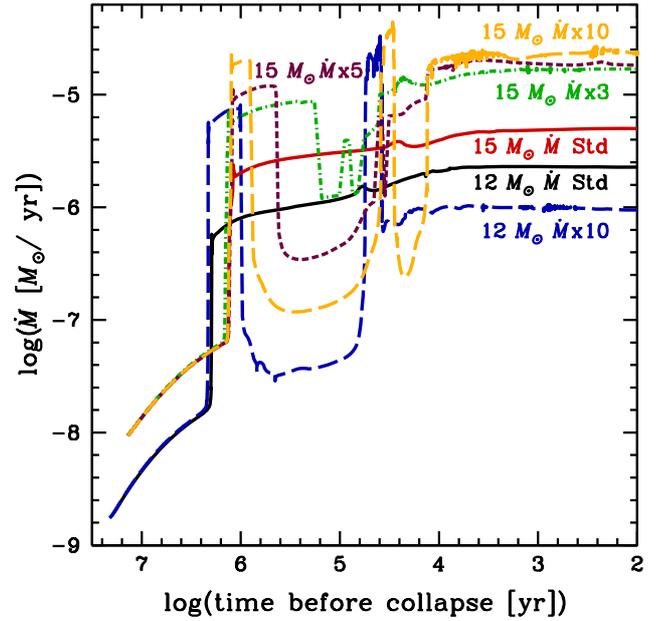}
\caption{Mass-loss rates of the models as a function of time, for the last million years of the stellar life. The labels near the curves indicate the corresponding model.}
\label{FigMdot}
\end{figure}

\begin{figure}
\centering
\includegraphics[width=.45\textwidth]{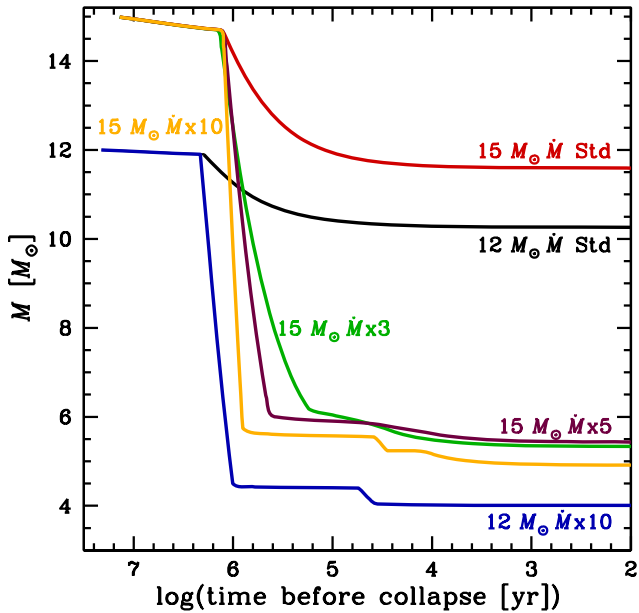}
\caption{Total mass of the models as a function of time. As a comparison, the final mass of the CO-cores are $2.06$ for model 12Std and $2.12\, M_{\sun}$ for 12x10, and $2.87$, $3.01$, $2.67$, and $2.74\, M_{\sun}$ for the models 15Std, 15x3, 15x5, and 15x10, respectively.}
\label{FigMass}
\end{figure}

The mass-loss rates of the models are shown in Fig.~\ref{FigMdot}, and the time evolution of their total mass in Fig.~\ref{FigMass}. The models with a standard mass-loss prescription show a strong increase ($1.5$ orders of magnitude) in the mass-loss rates as the star enters the RSG phase compared to the mass-loss rate during the MS. The models with increased mass loss reach strong mass-loss rates of around $10^{-5}\, [M_{\sun}\,\text{yr}^{-1}]$ when they enter the RSG phase. During that phase, the mass decreases quickly. The models with increased mass loss then enter the blue loop, during which the mass-loss rates are lower. Because the duration of the loop is longer for stars with stronger mass-loss rates (up to $\sim 50\%$ of the total duration of the central helium burning for the models with mass-loss rates increased by a factor of $10$), the final masses of the models with increased mass loss are comparable.

% SECTION 3 %%%%%%%%%%%%%%%%%%%%%%%%%%%%%%
\section{Supernova progenitors}\label{SecCompa}

\subsection{Comparison with observed YSG supernova progenitors}

In the right-hand panel of Fig.~\ref{FigHRD}, we show a zoom in the HRD, centred on the region where the end points of the evolutionary tracks lie. We also indicate the observed position of four recently discovered YSG SN progenitors: 1993J, 2008cn, 2009kr, and 2011dh. All the suspected progenitors have a metallicity near solar, except SN 1993J, whose metallicity is estimated at roughly twice solar \citep{Smartt2002a}. Also these SNe were of the quite rare types IIL or IIb (except SN 2008cn, classified as a type IIP SN). Classical evolutionary tracks of single star are unable to explain the position of these stars in the yellow area of the HRD. The binary channel is usually evoked to solve this problem \citep{Podsiadlowski1992a,Podsiadlowski1993a,Stancliffe2009a,Claeys2011a}. However, in at least one case (SN 2011dh), \citet{Maund2011a} indicate that no companion to the observed progenitor is detectable, making it more probable that this progenitor is a single star\footnote{\footnotesize{This does not mean that this star was not a binary in the past. According to \citet{Eldridge2011a}, secondaries can be at the origin of up to $15\%$ of all type IIb, IIL and IIn SNe.}}.

Looking at the end points of the tracks of the models with increased mass-loss rate, we see that they are in excellent positions to explain these YSG progenitors, covering the same $T_\text{eff}$ range as the observations. The progenitors of SN 2011dh and 2009kr could thus be a $15\, M_{\sun}$ star having encountered a stronger mass-loss rate during the RSG phase than currently adopted in stellar evolution codes. The slightly higher luminosity of the progenitors of SN 1993J and 2008cn indicates that they could have a slightly higher initial mass than our $15\, M_{\sun}$. As already mentioned in \citet{Smartt2009a}, the final luminosity of our models are roughly independent of the post-MS mass-loss history. Models with similar initial mass and rotation velocity end within a $\pm 0.1\,\text{dex}$ luminosity range\footnote{\footnotesize{However, varying the initial rotation velocity also leads to a similar dispersion \citep[see][]{Ekstrom2011a}.}}. The initial mass of an SN progenitor can thus be estimated using the luminosity of the progenitor alone.

% FIGURE Mass &&&&&&

\begin{figure}
\centering
\includegraphics[width=.45\textwidth]{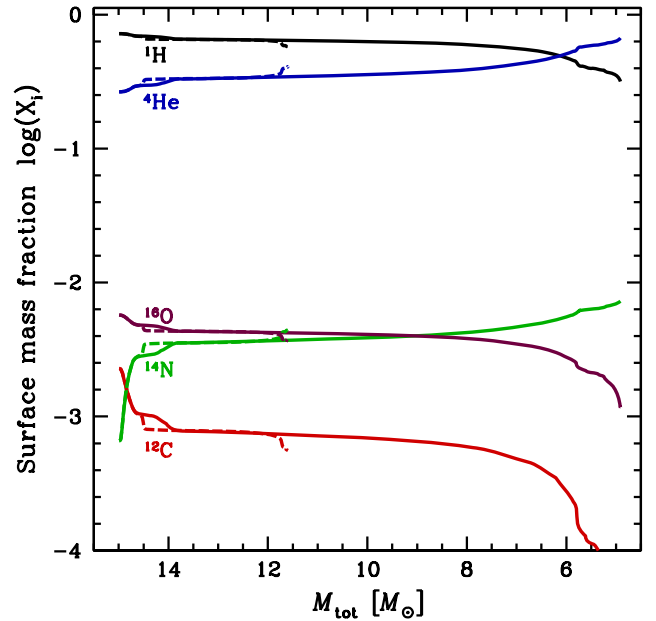}
\caption{Logarithm of the mass fraction of $^1\text{H}$, $^4\text{He}$, $^{12}\text{C}$, $^{14}\text{N}$, and $^{16}\text{O}$ at the surface as a function of the actual mass of the star, for the model 15Std (dashed line) and 15x10 (solid line).}
\label{FigAbund}
\end{figure}

The surface abundances of the main chemical elements as a function of the actual mass of the star are shown in Fig.~\ref{FigAbund}. For the model with standard mass-loss rates, we see that the surface chemical composition progressively evolves, revealing the products of H-burning through the CNO-cycle: a decrease in the H, C, and O abundances and an increase in the He and N abundances. In the case of an increase in the RSG mass-loss rate by a factor of ten, the same trend is observed, but much stronger. In that case, the surface H abundance is strongly depleted, becoming smaller than the He abundance. Measurements of the surface abundances indicating such trends would be a clue to strong mass loss during the RSG phase.

As three of the four observed SNe discussed in this paper are of type IIL or IIb, it is also important that our rotating single star models are compatible with such an explosion. As discussed in \citet{Heger2003a}, determining the supernova type on the basis of stellar models alone is not simple, as it depends strongly on the explosion properties. However, the absence of a plateau in the lightcurves of the types IIL and IIb SNe indicates that the hydrogen envelope ejected during the explosion must be small. We adopt here the same criterion as in \citet{Heger2003a} and \citet{Eldridge2004a} to distinguish between type IIP and IIL SNe: it is a type IIL if the ejected mass of hydrogen is lower than $2\, M_{\sun}$, otherwise it is a type IIP.

The ejected hydrogen amounts during the explosion are $3.8\, M_{\sun}$, $0.11\, M_{\sun}$, and $0.03\, M_{\sun}$ for the models 15Std, 15x3, and 15x10 respectively. The model with standard mass-loss prescription will thus produce a type IIP SN, and the other two a type IIL / IIb SN. An increased mass-loss rate during the RSG phase produces rotating single-star models that fit both the position in the HRD and the SN type simultaneously for the cases we consider in this paper. Also the quite low hydrogen content of the models with increased mass loss at the end of their lives could explain the lack of type IIP SNe from high-mass RSG \citep{Smartt2009b}.

We can wonder to what extent an increased mass-loss rate is usual during the RSG phase. As already mentioned, there are theoretical and observational arguments for \textit{and} against an increased mass-loss rate, making it very uncertain. In the right-hand panel of Fig.~\ref{FigHRD}, the positions of Galactic RSG are indicated \citep[small red circles,][]{Levesque2005a}. The time spent on the RSG branch depends on the mass-loss rate adopted. It corresponds to the whole central He-burning for the standard mass-loss prescription and to approximatively half of the same time for the models with mass loss increased by a factor of ten, before the star evolves in the blue part of the HRD. Models with increased mass mass loss are thus able to reproduce the location of the RSG-branch, as well as the models with standard mass loss \citep[see also][]{Ekstrom2011a}. However, that the identified YSG SN progenitors exhibit a rather large dispersion in $T_\text{eff}$, and on the other hand, there are a few type IIP SNe arising from stars with an initial mass between $\sim 15-18\, M_{\sun}$, which are probably RSG \citep{Smartt2009b}, indicates that ``standard'' and ``increased'' mass-loss rates RSG coexist, leading to the observed variety of type II SNe and their progenitors.

A way to determine the relative importance of each type of evolution would be to compare the relative number of type IIL/b SNe and type IIP SNe arising from stars with roughly the same initial mass. The actual data \citep[\textit{e.g.}][]{Smartt2009a} seem indicate that an increased mass-loss rate is relatively infrequent, due to the smaller number of type IIL/b SNe than type IIP. However, most of type IIP SNe arise from stars with initial mass lower than the ones considered in the present paper. If for some reason, an increased mass-loss rate is favoured for more massive stars \citep[as expected, for example, in the case of the pulsation-driven winds discussed in][]{Yoon2010b}, the relative contribution of both mass-loss scenarios should depend on the initial mass, and the physical processes leading to the increased mass-loss rate could be more effective for stars of around $15\, M_{\sun}$ than for smaller ones.

% FIGURE Abundances &&&&&&

One can also wonder whether the strong mass loss encountered by our models leave observable imprints in the SN spectra. The stellar winds during the RSG phase are slow winds, and might interact with the shock wave or the ejecta during the SN explosion. \citet{Filippenko1997a} mentions that type IIn SNe are SNe with a spectrum showing a strong interaction with a dense circumstellar medium, which indicates a strong mass-loss episode before the SN explosion. We cannot determine whether this should be the case here on the basis of our models. However, it would be interesting to find a observable signature of a strong mass loss in the SNe spectra.

\subsection{Low-luminosity Wolf-Rayet stars}

A particularly challenging question for the evolution of massive star is the existence of low-luminosity WC stars. They are not explained by current rotating single-star models \citep[][submitted]{Georgy2011b}, which produce much more luminous WC models. In this previous paper, we pointed out three solutions to this problem:
\begin{itemize}
\item an increased mass loss during some part of the evolution of the most massive stars (during the MS and/or during the previous WR phases),
\item an increased mass loss during the RSG phase of stars around $15$ - $20\, M_{\sun}$ (the case explored in this work),
\item close binary evolution, where some mass transfer occurs between the two components, allowing for the primary to lose its H-rich envelope. The effect on the evolution is similar to the previous item, instead that the mass loss is due to binary interactions \citep{Eldridge2008a}.
\end{itemize}

The calculations performed in this work indicate that, even with a strong mass loss during the RSG phase, our models are still far from becoming WR stars. Even if the surface abundances show a depletion of hydrogen and an increase in helium, they do not fulfil the criterion needed to be classified as a WR star \citep[see][]{Georgy2009a}. Moreover, it would be necessary to completely remove the H-rich envelope of the star to become a WC star. In view of the structure of our models at the end of the stellar lifetime, it seems very unlikely that low-luminosity WC stars can arise from stars in the $15$ - $20\, M_{\sun}$ mass range through the single-star channel.

% SECTION 4 %%%%%%%%%%%%%%%%%%%%%%%%%%%%%%
\section{Conclusions}\label{SecConclu}

Motivated by observational and theoretical indications that the current prescriptions for the mass-loss rate of RSG could be underestimated and by the increasing number of YSG SN progenitors, we performed a set of simulations of the evolution of rotating stars between $12$ and $15\, M_{\sun}$, with various mass-loss rates during the RSG phase.

We show that these kinds of models are able to reproduce the positions of the YSG progenitors in the HRD, fitting the effective temperature well where these progenitors are observed, as well as the characteristics of the ejecta, leading to type IIL or IIb SNe. Even if the exact physical mechanism leading to these high mass-loss rates still has to be determined, the consequent stellar evolution could thus be an alternative to the binary channel, which is also able to reproduce the observed characteristics of the progenitor.

Moreover, the low hydrogen content of the YSG progenitor will favour an explosion as a type IIL/IIb SN, giving a possible explanation for the lack of massive RSG as progenitors of type IIP SNe. A more exhaustive study of the consequences of such an increased mass-loss rate will be addressed in a future paper, with a more extended set of mass, rotational velocities, and metallicities.

Finally, the observed dispersion in effective temperature of the identified YSG progenitors, as well as the occurrence of massive type IIP SNe, indicates that all the stars do not undergo the same mass-loss rate increase. However, a set of models with both various enhancements of the mass-loss rates and standard mass-loss rate is able to reproduce the whole range of observed progenitors.

% ACKNOWLEDGEMENTS %%%%%%%%%%%%%%%%%%%%%%%%%%%%%%
\begin{acknowledgements}
The author would like to thank the referee John J. Eldridge for his precious comments that considerably improved this work.
\end{acknowledgements}

% BIBLIOGRAPHIE %%%%%%%%%%%%%%%%%%%%%%%%%%%%%%
\bibliographystyle{aa}
\bibliography{MyBiblio}
%%%%%%%%%%%%%%%%%%%%%%%%%%%%%%%%%%%%%%%%

\end{document}